# Decomposing 3D Neuroimaging into 2+1D Processing for Schizophrenia Recognition

Mengjiao Hu[a,b], Xudong Jiang[c,*], Kang Sim[d,e], Juan Helen Zhou[b,f], Cuntai Guan[g]

[a]*Interdisciplinary Graduate Programme, NTU Institute for Health Technologies, Nanyang Technological University, Singapore*
[b]*Center for Sleep and Cognition, Department of Medicine, Yong Loo Lin School of Medicine, National University of Singapore, Singapore*
[c]*School of Electrical & Electronic Engineering, Nanyang Technological University, Singapore*
[d] *West Region, Institute of Mental Health (IMH), Singapore*
[e]*Department of Research, Institute of Mental Health (IMH), Singapore*
[f] *Translational Magnetic Resonance Research, Yong Loo Lin School of Medicine, National University of Singapore, Singapore*
[g]*School of Computer Science and Engineering, School of Electrical and Electronic Engineering, Nanyang Technological University, Singapore*

**Abstract**

Deep learning has been successfully applied to recognizing both natural images and medical images. However, there remains a gap in recognizing 3D neuroimaging data, especially for psychiatric diseases such as schizophrenia and depression that have no visible alteration in specific slices. In this study, we propose to process the 3D data by a 2+1D framework so that we can exploit the powerful deep 2D Convolutional Neural Network (CNN) networks pre-trained on the huge ImageNet dataset for 3D neuroimaging recognition. Specifically, 3D volumes of Magnetic Resonance Imaging (MRI) metrics (grey matter, white matter, and cerebrospinal fluid) are decomposed to 2D slices according to neighboring voxel positions and inputted to 2D CNN models pre-trained on the ImageNet to extract feature maps from three views (axial, coronal, and sagittal). Global pooling is applied to remove redundant information as the activation patterns are sparsely distributed over feature maps. Channel-wise and slice-wise convolutions are proposed to aggregate the contextual information in the third view dimension unprocessed by the 2D CNN model. Multi-metric and multi-view information are fused for final prediction. Our approach outperforms handcrafted feature-based machine learning, deep feature approach with a support vector machine (SVM) classifier and 3D CNN models trained from scratch with better cross-validation results on publicly available Northwestern University Schizophrenia Dataset and the results are replicated on another independent dataset.

---

*Corresponding author. Tel.: +65 67905018
*Email address*: exdjiang@ntu.edu.sg (Xudong Jiang)

## 1. Introduction

Increasing numbers of populations suffer from chronic psychiatric disorders such as schizophrenia and depression, causing great burdens to patients' families and society [1,2]. Yet, the diagnosis of such psychiatric disorders is typically based on clinical interviews and clinical systems, which could be challenged by the complex symptoms and multiple confounders such as race, gender, and medication effects [3,4].

Machine learning approaches demonstrated remarkable potential on neuroimaging recognition tasks. Previous work on computer-aided classification mainly focused on handcrafted feature-based machine learning approach, which requires feature extraction and reduction before classification. Support vector machine (SVM) has been widely used in classification of schizophrenia patients and healthy controls using Magnetic Resonance

Imaging (MRI) features such as voxel-based morphometry and cortical thickness [5–8]. Features with high dimensionality are often processed with Principal Component Analysis (PCA) or knowledge-based ROI selection for dimensionality reduction. The performance of handcrafted feature-based machine learning varies among different datasets across studies and more importantly, the feature extraction and selection prior to classification might be non-generalizable when challenged by confounders such as disease duration, race, medication effect, and so on [9].

Deep learning approaches such as Convolutional Neural Network (CNN) has been successfully applied on medical images for the classification of brain tumor [10], breast cancer [11], Alzheimer's disease [12,13] and so on. CNN is capable of feature learning, which is particularly important for psychiatric disorders such as schizophrenia that have complex and widely distributed brain alterations [9,14]. The deep model architecture and nonlinear layers of CNN allow investigating the complicated data patterns of images [15]. Ensemble methods such as gating and fusion further improves the performance of CNN on medical image classification tasks [16,17]. Nevertheless, the application of deep learning to classify psychiatric disorders is restricted due to the lack of representative slices and small sample size. There are no specific slices indicating differences between patients and healthy controls, which limits the training of 2D CNN models with slices. Applications of powerful pre-trained 2D CNN models to medical images are also limited to specific 2D slices or regions of interest [18,19]. Due to the lack of available powerful networks pre-trained on medical images, most studies utilized models pre-trained with ImageNet [20]. Finetune methods which tune the pre-trained model parameters for new classification tasks are generally applied on relatively large datasets due to the requirement of re-training [21,22]. Deep feature methods classify the deep features extracted from the frozen pre-trained networks with fully-connected layers or conventional classifiers such as SVM and logistic regression (LR) and can be applied on relatively small datasets [20,23]. Extra feature selection via pooling layers [24,25], or statistical methods [26] were employed to reduce the high dimensionality of extracted features. Most studies reported superior or comparable performance comparing to handcrafted feature-based machine learning. Yet most of the studies were slice or ROI-based, in which specific slices or ROIs were selected to re-train the 2D networks or to extract features from 2D networks, limiting the application of such methods to psychiatric disorders which has no representative slices. Recent works on whole-brain MRI images utilized 3D CNN models for classifying schizophrenia patients and healthy controls [27–29]. Yet, challenged by the high demand of computational cost and training sample size, the architecture of 3D CNN models utilized remains shallow and therefore may not be able to extract the complicated data patterns. To exploit the powerful 2D deep CNN pre-trained on huge database such as ImageNet, we propose a framework that decomposes 3D neuroimages to 2+1D processing for the classification of schizophrenia patients and healthy controls. Specifically, 3D MRI images are decomposed to 2D slices with regard to neighboring voxel positions to resemble the characteristics of natural images. Firstly, feature maps are extracted from 2D slices of MRI metrics via a pre-trained 2D CNN model from three views (axial, coronal, and sagittal) and max-pooled according to decomposition positions. Secondly, learning-based global pooling along the view dimension reduces the number of slices and channels to remove redundant information as the activation patterns are sparsely distributed over feature maps. Thirdly, the pooled feature maps are used to train an 1D network using channel-wise convolution and slice-wise convolution to aggregate the contextual information in the unprocessed view dimension. Multi-view and multi-metric information are cooperated for the final prediction. Our approach overperforms handcrafted feature-based machine learning, 2D CNN models as well as 3D CNN models with better classification results and lower computational cost on publicly available Northwestern University Schizophrenia Dataset and the results are further validated on an independent dataset.

The paper is structured as follows: Section 2 presents the formulations and optimization procedure of the proposed 2+1D framework. Section 3 describes the data used and the data processing procedures. Section 4 illustrates the experimental results, discusses the performance of our approach and previous methods, and verifies the effect of the proposed methods using ablation study. We conclude in Section 5.

## 2. Methods

The proposed framework is presented in detail in this section. The overall flowchart is presented in Fig. 1. The selected pre-trained 2D CNN model and feature map extraction procedure with image decomposition and feature max-pooling are introduced first, followed by global pooling in slice and channel dimensions. Then the classification network consisting of channel-wise and slice-wise convolution is presented. Lastly, multi-view and multi-metric fusion is introduced.

**2.1 Image decomposition, feature extraction and 3D max-pooling**

75

The 3D MRI volumes are decomposed into 2D slices for feature extraction via a pre-trained 2D network. To stimulate the characteristics of natural images for effective and accurate feature extraction, the decomposition involves not only slice separation, but also 3D down-sampling according to neighboring voxel locations. Each 3D MRI volume is down-sampled into eight volumes and then further separated as slices from axial, coronal and sagittal views. The 8 down-sampled MRI volumes share similar contextual information and the resulting slices have sharpened boundaries that resemble the attributes of natural images so that they better fit the 2D network trained by natural images.

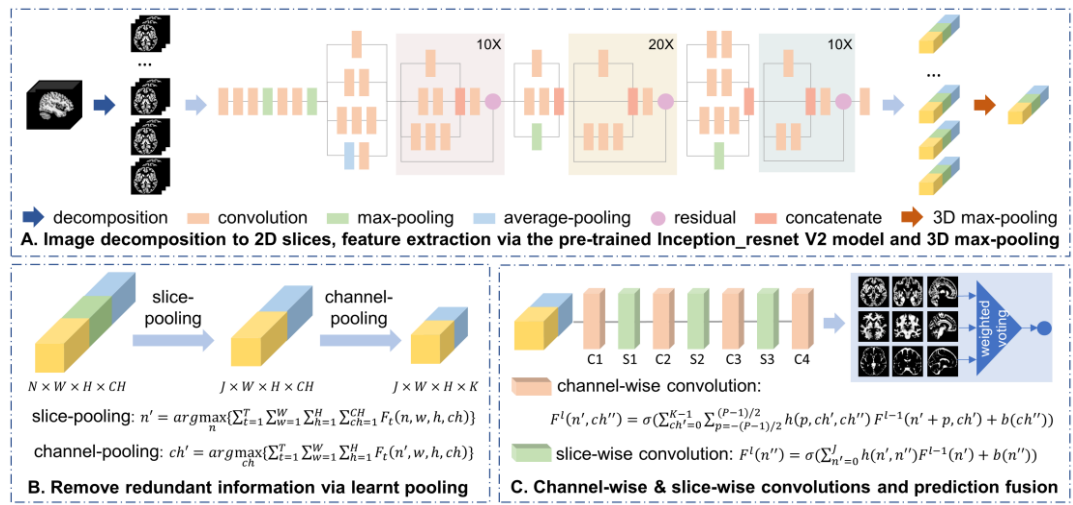

Fig. 1. Framework of the proposed method. Step (A) decomposes 3D MRI images into 2D slices regarding neighboring positions, extracts feature maps from the Inception_resnet V2 model pre-trained on ImageNet dataset, and performs 3D max-pooling on feature maps to retain the max value among the 8 spatially nearest neighbors. Slice-pooling and channel-pooling in step (B) removes the redundant information via learning-based global pooling on the slice and channel dimensions. Step (C) trains a 1D CNN model to perform channel-wise and slice-wise convolution to learn, extract and integrate the information on the view dimension unprocessed by the 2D CNN. Prediction fusion in step (C) merges the information from multi brain metrics and multi views for the final prediction.

The pre-trained model Inception_resnet V2 is adopted in this study for feature extraction. Inception_resnet V2 is one of the most powerful deep learning models for image classification tasks [30]. It utilizes inception blocks which allows multi-level feature extraction at the same time and residual connections which addresses the degradation problem for training very deep networks. The architecture of Inception_resnet V2 is shown in Fig. 1A.

Passing a slice through the pre-trained Inception_resnet V2 model in a forward manner, we get a feature map at the last convolutional layer. For eight down-sampled volumes, we obtained eight sets of extracted feature maps $F^k(n, w, h, ch)$, $k = 1, 2, \ldots, 8$. Max-pooling is performed across eight sets of feature maps as in equation (1) to extract the maximum values since the down-sampled volumes are spatially 8 nearest neighbors that share similar information with close voxel coordinates in the original 3D volume. The max-pooling over the 8 corresponding feature maps also makes the retained feature insensitive to the spatial location.

$$F(n, w, h, ch) = \max_k \{F^k(n, w, h, ch)\} \tag{1}$$

where $F(n, w, h, ch)$ represents the set of feature maps that have maximum values pooled among eight sets of feature maps $F^k(n, w, h, ch)$ with element-wise values, $n$ represents the slice index, $n \in [1, N]$, $N = 61$ for axial and sagittal views, $N = 73$ for coronal view, $w, h, ch$ represent the width, height and channel indices of the feature maps respectively, $w \in [1, W]$, $h \in [1, H]$, $ch \in [1, CH]$, $W = 3, H = 3, CH = 1536$.

All 3D MRI images are now converted into feature maps $F(n, w, h, ch)$ after image decomposition, feature extraction by the pre-trained model and 3D max-pooling. In the following section 2.2 and 2.3, we will further process them along the slice dimension (view axis $n$) and channel dimension (channel axis $ch$) to remove redundant information and learning new features from the unprocessed dimension.

## 2.2 Remove redundant information via learning-based global pooling in the slice and channel dimensions

The obtained feature maps $F(n, w, h, ch)$ has a large number of slices as there is no feature extraction and pooling operation performed on the view axis by the 2D CNN model. Thus, some slices may not contain any discriminative information as activation patterns contain sparsely distributed information. Meanwhile, the width and height of the feature maps in the last convolutional layer are small while the number of channels is large. In our case, the Inception_resnet V2 model extracts feature maps with dimension $3 \times 3 \times 1536$ at the last convolutional layer. As the 2D CNN model is pre-trained by natural images, many channels of feature maps may not contain any discriminative information for MRI slices. Therefore, it is critical to remove the redundant information in both slice and channel dimensions before further learning the discriminative information along the third dimension unprocessed by the 2D CNN model.

We propose to perform global pooling in the slice dimension via equation (2). The values of the feature maps are summed up along width, height, and channel dimension to obtain a representative value for each slice. The rectified linear units (ReLU) activation is utilized in the pre-trained network and hence higher values in extracted feature maps correspond to higher responses to structures that are more important for classification. Slices from different positions of the 3D image cover different parts of the brain. Some slices may contain mostly background which results in small values in extracted feature maps. The redundant information is removed by discarding corresponding feature map of slices. Furthermore, we sum up the values over all MRI training data to avoid overfitting to specific individuals. We found $J$ desired slice indices $n'$ via equation (2) where $J = N/2$. The pooled feature map is represented by $F(n', w, h, ch)$ with updated slice indices.

$$n' = arg\max_{n}\{\sum_{t=1}^{T} \sum_{w=1}^{W} \sum_{h=1}^{H} \sum_{ch=1}^{CH} F_t(n, w, h, ch)\} \quad (2)$$

where $F_t(n, w, h, ch)$ represents the max-pooled feature maps obtained from equation (1) for subject $t$ with element-wise values. $W, H, CH$ represent the width, height, and channel, respectively. $T$ represents the number of subjects (samples) in training data. The desired slice indices $n'$ learned from MRI training data are then used for all MRI data for further learning and prediction. Global pooling in the channel dimension is proposed via equation (3). The feature maps are summed along width and height dimension to form a feature cube for each channel in each slice. The channels of small feature value on the MRI training data are discarded. For each slice, we found $K$ desired channel indices $ch'$ via equation (3) where $K = CH/4$ is selected. The new feature map is represented by $F(n', w, h, ch')$ with updated channel indices.

$$ch' = arg\max_{ch}\{\sum_{t=1}^{T} \sum_{w=1}^{W} \sum_{h=1}^{H} F_t(n', w, h, ch)\} \quad (3)$$

where $F_t(n', w, h, ch)$ represents the updated feature maps obtained from equation (2) for each subject $t$ with element-wise values. $W$, $H$ represent the width and height, respectively. $T$ represents the number of subjects (samples) in training data. The desired channel indices $n'$ learned from MRI training data are then used for all MRI data for further learning and prediction.

**2.3 Channel-wise & slice-wise convolutions and multi-brain metric**

### & multi-view fusion

To learn, integrate and extract the information on the view axis which has not been processed by the pre-trained 2D CNN model, two types of convolutions are proposed to perform alternately. Following equation (4), channel-wise convolution performs $(P,1,1)$ convolution in the slice, width and height dimension, and the number of channels is updated according to layers. With a small number of nodes chosen at the bottom layer, the dimensionality of channels is further reduced. The convolutional kernel $(P,1,1)$ extracts contextual information from the slice dimension without interrupting the other two dimensions that have been well-processed by the 2D network.

$$F^l(n',ch'') = \sigma(\sum_{ch'=0}^{K-1} \sum_{p=-\frac{P-1}{2}}^{\frac{P-1}{2}} h(p,ch',ch'') F^{l-1}(n'+p,ch') + b(ch'')) \quad (4)$$

where $F^l$ and $F^{l-1}$ represents the feature map in the $l$-th layer and the previous layer. For symbolic simplicity, we omit indices $w$ and $h$ of $F(n',w,h,ch')$. $F^l(n',ch'')$ and $F^{l-1}(n'+p,ch')$ represent their element-wise value with slice and channel dimension coordinates. $h(\cdot)$ represents the weights of convolutional kernel connecting $F^l$ and $F^{l-1}$, $b(\cdot)$ represents the bias term, $\sigma(\cdot)$ represents the nonlinear activation function, i.e., the ReLu activation function, $(\sigma(a) = \max(0, a))$, $K$ represents the total number of channels in the input feature map, $P$ represents the size of the convolutional kernel.

Table 1. Channel-wise and slice-wise convolutional neural network architecture.

| Layer | Feature map dimension | Kernel size | Num. of nodes |
|---|---|---|---|
| C1 | $J \times W \times H \times K$ | (3,1,1) | 32 |
| S1 | $J \times W \times H \times 32$ | (1,1,1) | $J/2$ |
| C2 | $J/2 \times W \times H \times 32$ | (3,1,1) | 32 |
| S2 | $J/2 \times W \times H \times 32$ | (1,1,1) | $J/4$ |
| C3 | $J/4 \times W \times H \times 32$ | (3,1,1) | 64 |
| S3 | $J/4 \times W \times H \times 64$ | (1,1,1) | $J/8$ |
| C4 | $J/8 \times W \times H \times 64$ | (1,1,1) | 128 |
| FC | $128WH(J/8)$ | - | 2 |

Equation (5) demonstrates the slice-wise convolution which performs (1,1,1) convolution in width, height and channel dimension while updating the number of slices by changing the number of nodes.

$$F^l(n'') = \sigma\left(\sum_{n'=0}^{J-1} h(n', n'') F^{l-1}(n') + b(n'')\right) \quad (5)$$

where $F^l$ and $F^{l-1}$ represents the feature map in the $l$-th layer and the previous layer. For symbolic simplicity, we omit indices $w, h$ and $ch''$ of $F(n', w, h, ch'')$. $F^l(n'')$ and $F^{l-1}(n')$ represent their element-wise value with slice dimension coordinates. $J$ represents the total number of slices in the $(l-1)$-th layer. $h$, $b$ and $\sigma$ have the same meaning as those in the equation (4).

The proposed 1D channel and slice-wise alternate convolution network is trained by the MRI training data. The architecture of the proposed network is illustrated in Fig. 1C and the input feature map dimension, kernel size, number of nodes in each layer are listed in Table 1. The feature map obtained after image decomposition, feature extraction and 3D max-pooling in step A has dimension $N \times W \times H \times CH$, where $N = 61$ for axial and sagittal views, $N = 73$ for coronal view, $W = 3, H = 3, CH = 1536$. After global pooling in slice and channel dimension, the feature map has dimension $K \times W \times H \times J$, where $J = N/2, K = CH/4$. Channel-wise convolution performs (3,1,1) convolution with an increasing number of nodes to extract deep features whereas slice-wise convolution reduces the dimensionality of slice dimension by (1,1,1) convolutions with reducing number of nodes. The dimensionality of slice finally becomes comparable to the width and height comparable to width and height which were processed by the pre-trained 2D deep network. The fully-connected (FC) layer takes a vectorized input from the previous layer and generates two outputs as classification results.

The input 3D images include three brain metrics: gray matter, white matter, and cerebrospinal fluid maps. Three brain metrics provide information from different perspectives of the brain structures. The 2D slices were decomposed from three views: axial, coronal, and sagittal. Therefore, the final prediction is determined by weighted voting of multi-brain metric and multi-view results.

## 3. Experimental Settings

### 3.1 Datasets

Two independent MRI datasets of schizophrenia and controls which were comparable for age and gender for both groups were used in this study (Table 2). The Northwestern University Schizophrenia Data and Software Tool (NUSDAST) is a repository of schizophrenia neuroimaging data collected from over 450 schizophrenia patients and healthy controls, which is publicly available on SchizConnect platform [31,32]. Overall, 141 schizophrenia patients and 134 healthy controls from this public dataset were included after quality control. Another dataset from the Institute of Mental Health (IMH), Singapore, were included as an independent dataset [33,34]. In this dataset, 148

Table 2. Demographics of selected subjects.

|  | NUSDAST | | IMH | |
|---|---|---|---|---|
|  | SZ | HC | SZ | HC |
| Subject number | 141 | 134 | 148 | 76 |
| Age(years) mean (SD) | 35.1 (12.8) | 32.9 (14.0) | 32.7 (9.0) | 31.3 (9.8) |
| Sex (M/F) | 90/51 | 72/62 | 102/46 | 47/29 |

Abbreviations: SZ - Schizophrenia Patients, HC - Healthy Controls, SD - standard deviation, M – Male, F - Female

schizophrenia and 76 healthy controls were included after quality control.

**3.2 Preprocessing**

Processing of the structural MRI data was performed using Computational Anatomy Toolbox in Statistical Parametric Mapping 12 (SPM12) for voxel-wise estimation of grey matter, white matter, and cerebrospinal fluid compartments [35]. Images with motion artifacts were excluded after visual quality control. Subject-level probability maps were obtained with the following steps: (i) skull stripping; (ii) linear (FLIRT) and nonlinear (FNIRT) registration to the Montreal Neurological Institute (MNI) 152 standard space [36]; (iii) segmentation of the brain into gray matter, white matter, and cerebrospinal fluid compartments with 1.5 mm isotropic resolution ; (iv) modulation by multiplying voxel values with the linear and nonlinear components of the Jacobian determinant.

**3.3 Model training and performance evaluation**

5-fold cross-validation was utilized for multiple runs of training and performance evaluation, which averaged the test error over multiple train–test splits. Results were reported based on the average of ten repeats of 5-fold cross-validation to reduce randomness caused by the data and system. A random seed was assigned for data splitting to ensure that the same data split was used in each round. Stochastic gradient descent (SGD) optimizer and Binary_crossentropy loss were used for model training. Model was trained with a NVIDIA Tesla V100 Tensor Core and each epoch took 1 second. The model performance is quantitatively evaluated via accuracy, sensitivity, specificity, and area under the receiver operating characteristic curve (ROC-AUC).

# 4. Results

The classification results in previous studies vary due to different data selection and exclusion criteria, preprocessing protocols, and model implementation procedure. Therefore, we performed comparisons among handcrafted feature-based machine learning, 2D CNN models using deep feature approach, 3D CNN models trained from scratch and our 2+1D approach on the same datasets in this study for fair comparisons.

**4.1 Compare with handcrafted feature-based machine learning**

According to the literature, voxel-based morphometry is one of the most widely utilized handcrafted feature and SVM is the most widely applied classifier. To compare with the state of art approach using handcrafted feature-based machine learning, we used the grey matter, white matter, and cerebrospinal fluid probability maps extracted by SPM12 toolkit and trained both linear and nonlinear SVM for classification. The classification results are presented in Fig. 2 and Table 3. Our approach greatly outperforms handcrafted feature-based machine learning with a ~15% accuracy gap and 0.16 ROC-AUC gap on the NUSDAST dataset. Similar results were replicated on the IMH dataset. Handcrafted feature-based machine learning approaches that extract features prior to the classification might not be sufficiently discriminative, especially for schizophrenia which has subtle, mixed and widespread brain anatomical changes [7, 36]. On the other hand, CNN automatically extract features from the contextual information which mitigates the subjectivity and variability in selecting relevant features and hence leading to the superior performance.

Table 3. Performance comparisons with previous methods on the NUSDAST and IMH datasets.

| Method | NUSDAST results | | | | IMH results | | | | #Parameters | #Layers |
|---|---|---|---|---|---|---|---|---|---|---|
| | Acc(%) | Sp(%) | Se(%) | ROC-AUC | Acc(%) | Sp(%) | Se(%) | ROC-AUC | | |
| **Handcrafted Feature-based Machine Learning** | | | | | | | | | | |
| Linear SVM | 69.81 | 65.61 | 73.77 | 0.734 | 71.80 | 40.75 | 87.79 | 0.787 | - | - |
| Nonlinear SVM | 66.50 | 68.58 | 64.51 | 0.713 | 69.63 | 51.42 | 79.13 | 0.712 | - | - |
| **2D CNN model** | | | | | | | | | | |
| Deep feature + SVM | 68.35 | 63.45 | 73.05 | 0.738 | 76.32 | 48.75 | 90.51 | 0.835 | - | - |
| **3D CNN model** | | | | | | | | | | |
| 3D CNN trained from scratch | 79.27 | 80.44 | 78.15 | 0.813 | 80.17 | 65.12 | 87.81 | 0.825 | 729,058 | 19 |
| **Our approach** | 84.30 | 84.54 | 84.09 | 0.896 | 82.65 | 76.49 | 85.86 | 0.894 | 115,259 | (571)+8 |

Abbreviations: Acc - Accuracy, Sp – Specificity, Se - Sensitivity, ROC-AUC - area under the receiver operating characteristic curve

## 4.2 Compare with 2D CNN models

2D CNN models trained from scratch were not available due to the lack of representative slices in schizophrenia and disconvergence caused by confusing labels when using whole-brain data. Therefore, we implemented a deep feature approach combined with SVM as the benchmark for approaches utilizing 2D CNN models.

Deep feature approach utilizes features extracted from pre-trained CNN models and performs classification with conventional classifiers such as SVM and LR, which provides a practical way for applying pre-trained 2D CNN models for the classification of schizophrenia patients and healthy controls using 3D MRI data. To compare with the benchmark deep feature approach, we implemented an SVM classifier to classify deep features. The feature maps extracted from the Inception resnet V2 model pre-trained on the ImageNet database for each 2D slice from each view (axial, coronal, and sagittal) and each brain metric (grey matter, white matter, and cerebrospinal fluid) were combined and flattened as large feature vector at the individual level. Linear SVM was trained to classify the feature vectors. The classification results are presented in Fig. 2 and Table 3. Deep feature approach takes advantage of networks pre-trained on large datasets and enables the feature extraction for small datasets [20]. Yet, the high dimensionality of the unprocessed view axis dimension and channel dimension limits the performance of deep feature approach. Our approach effectively removes the redundant information extracted from pre-trained 2D CNN that is irrelevant to the MRI data, extracts and integrates the information in the unprocessed view dimension via three steps, achieving ~16% improvement on accuracy and 0.083 ROC-AUC improvement than the deep feature + SVM approach on the NUSDAST dataset. Similar results were obtained on the independent IMH dataset.

**4.3 Compare with 3D CNN models**

3D CNN models operate convolution and pooling in a cubic manner with 3D feature volumes instead of 2D feature maps to fully utilize the contextual information of 3D images. However, the high computational cost and requirement of a large sample size for training prevent the development of very deep 3D CNN models and applications to datasets with a small sample size. There are also no available pre-trained 3D CNN models for medical imaging classification. In our previous work, we proposed 3D Naïve CNN with different architectures trained from scratch [26, 28]. The model that has the best performance employs the inception module and residual connection architectures. Multi-channel inputs including grey matter, white matter, and cerebrospinal fluid probability maps are used to provide additional information for classification. The classification results of the 3D CNN model with the best performance are presented in Fig. 2 and Table 3. Our approach obtained ~5% improvement on accuracy and much more balanced sensitivity and specificity. Note that the 3D CNN model trained from scratch consists of 19 3D layers while our approach utilized a pre-trained 571-2D-layer network and an 8-1D-layer network trained from scratch. The 3D CNN has 729,058 trainable parameters while our model has only 115,259 parameters. Another recent work trained an 8-layer sequential 3D CNN models and has 1,246,434 trainable parameters [27]. The training expenses increase exponentially when 3D CNN models get deeper and employs advanced architectures. The sample size required to train the model also increases greatly. Therefore, the current shallow 3D CNN models may not be able to fully extract the complicated data patterns. Our approach utilizes a very deep pre-trained 2D CNN model which has 571 layers to extract the feature maps from decomposed 2D slices and uncover the contextual information from the deep layers. As a result, our approach achieved higher accuracy than the 3D CNN models with fewer parameters and significantly lower computational cost. Our framework highlights the possibility of utilizing very deep 2D networks with low computational cost on datasets with a small sample size.

**4.4 Ablation study**

We performed ablation analysis step by step to verify the feasibility and usefulness of each step in our proposed framework, including image decomposition & 3D max-pooling, global pooling in the slice & channel dimensions, channel-wise & slice-wise alternate convolutions, and multi-metric & multi-view fusion. Results presented in Table 4 indicate that every step contributed significantly as all methods with an ablation have degraded performance. Results without image decomposition & 3D max-pooling were obtained using 2D slices directly split from the 3D volumes without down-sampling and hence no 3D max-pooling was performed. The ~9% accuracy reduction verifies our assumption that down-sampling according to neighboring voxel positions sharpened the boundaries of MRI images and resembled the characteristics of natural images, leading to better fitting in the 2D CNN model trained with the ImageNet dataset. Results without global pooling in slice & channel dimensions has a ~ 4% accuracy reduction and much higher computational cost as much more parameters were needed to train the 1D CNN model. The proposed 1D slice-wise & channel-wise alternate convolutions play the most important role as reflected by the ~13% accuracy reduction. Our results highlight the importance of engaging information in the view axis and removing redundant information. The improvement with multi-metric and multi-view fusion is also compelling with ~11% gap in accuracy, highlighting the critical role of multi-view processing of 3D images.

Table 4. Ablation study on the NUSDAST dataset.

| Method | Accuracy | Specificity | Sensitivity |
|---|---|---|---|
| No 1D slice-wise & cannel-wise alternate convolutions | 71.25% | 74.62% | 68.10% |
| No multi-metric & multi-view fusion | 72.84% | 66.80% | 78.56% |
| No decomposition & 3D max-pooling | 75.44% | 75.90% | 75.00% |
| No global pooling in slice & channel dimension | 79.84% | 80.84% | 78.86% |
| Our approach | 84.30% | 84.54% | 84.09% |

## 5. Conclusion

We have presented an efficient and effective framework that processes 3D neuroimaging data by 2+1D network for the classification of schizophrenia patients and healthy controls. We propose image decomposition and 3D maxpooling to resemble the characteristics of natural images for fully-utilization of models pre-trained on the huge natural image database, the ImageNet dataset. Global pooling in slice and channel dimension learned from MRI training data removes the redundant information that are irrelevant to MRI data without interrupting the feature map spatial architecture. Channel-wise and slice-wise convolutions further aggregate the contextual information in the view dimension unprocessed by the 2D network. Prediction fusion integrates multi-metric and multi-view information for the final prediction. Despite the promising results, there are a few limitations of this study. The sample size is rather small for network training involving convolutional layers, which might limit the performance of the model and result in low generalizability. The network explored in this study are pre-trained on a natural image database, the ImageNet dataset, which is very different from medical images. Future work may explore networks pre-trained on medical images which share more similarity with MRI images.

The proposed framework for 3D neuroimaging classification utilizes a deep 2D network pre-trained by huge natural image dataset ImageNet plus a 1D network trained by a small number of neuroimages. Our approach addresses the challenging classification problem of psychiatric disorders that have no representative slice as well as the challenge of limited number of training samples in medical imaging classification tasks. The framework decomposes 3D neuroimaging data into 2+1D and proposes new pooling and convolution strategies to remove redundant information and integrate deep contextual information of the data. The experimental results outperform handcrafted feature-based machine learning, 2D CNN and 3D CNN models with greatly reduced computational costs, highlighting the possibility of utilizing the powerful pre-trained deep 2D CNN models for 3D neuroimaging classification, especially future classification tasks for other psychiatric disorders and deep CNN models pre-trained medical images. This study provides an objective 3D neuroimaging-based framework for individual diagnosis of schizophrenia the lays the foundation for computer-aid diagnosis of psychiatric disorders utilizing deep learning approaches.

**Acknowledgment**

This work was supported by NTU Institute for Health Technologies, Interdisciplinary Graduate Programme, Nanyang Technological University and Yong Loo Lin School of Medicine, National University of Singapore.

**Conflict of interest**

The authors declare that they have no conflict of interest.

**References**

[1] A. Radaic, D. Martins-de-Souza, The state of the art of nanopsychiatry for schizophrenia diagnostics and treatment, Nanomedicine Nanotechnology, Biol. Med. 28 (2020) 102222. https://doi.org/10.1016/j.nano.2020.102222.

[2] H.Y. Chong, S.L. Teoh, D.B.C. Wu, S. Kotirum, C.F. Chiou, N.

Chaiyakunapruk, Global economic burden of schizophrenia: A systematic review, Neuropsychiatr. Dis. Treat. 12 (2016) 357–373. https://doi.org/10.2147/NDT.S96649.

[3] A.H. Fanous, R.L. Amdur, F.A. O'Neill, D. Walsh, K.S. Kendler, Concordance between chart review and structured interview assessments of schizophrenic symptoms, Compr. Psychiatry. 53 (2012) 275–279. https://doi.org/10.1016/j.comppsych.2011.04.006.

[4] J.L. Kennedy, C.A. Altar, D.L. Taylor, I. Degtiar, J.C. Hornberger, The social and economic burden of treatment-resistant schizophrenia: A systematic literature review, Int. Clin. Psychopharmacol. 29 (2014) 63–76. https://doi.org/10.1097/YIC.0b013e32836508e6.

[5] R. Chin, A.X. You, F. Meng, J. Zhou, K. Sim, Recognition of Schizophrenia with Regularized Support Vector Machine and Sequential Region of Interest Selection using Structural Magnetic Resonance Imaging, Sci. Rep. 8 (2018) 1–10. https://doi.org/10.1038/s41598-018-32290-9.

[6] S. Liang, Y. Li, Z. Zhang, X. Kong, Q. Wang, W. Deng, X. Li, L. Zhao, M. Li, Y. Meng, F. Huang, X. Ma, X.M. Li, A.J. Greenshaw, J. Shao, T. Li, Classification of first-episode schizophrenia using multimodal brain features: A combined structural and diffusion imaging study, Schizophr. Bull. 45 (2019) 591–599. https://doi.org/10.1093/schbul/sby091.

[7] J.L. Winterburn, A.N. Voineskos, G.A. Devenyi, E. Plitman, C. de la Fuente-Sandoval, N. Bhagwat, A. Graff-Guerrero, J. Knight, M.M. Chakravarty, Can we accurately classify schizophrenia patients from healthy controls using magnetic resonance imaging and machine learning? A multi-method and multi-dataset study, Schizophr. Res. 214

(2019) 3–10. https://doi.org/10.1016/j.schres.2017.11.038.

[8]  M. Nieuwenhuis, N.E.M. van Haren, H.E. Hulshoff Pol, W. Cahn, R.S. Kahn, H.G. Schnack, Classification of schizophrenia patients and healthy controls from structural MRI scans in two large independent samples, Neuroimage. 61 (2012) 606–612. https://doi.org/10.1016/j.neuroimage.2012.03.079.

[9]  M.R. Arbabshirani, S. Plis, J. Sui, V.D. Calhoun, Single subject prediction of brain disorders in neuroimaging: Promises and pitfalls, Neuroimage. 145 (2017) 137–165. https://doi.org/10.1016/j.neuroimage.2016.02.079.

[10] M.W. Nadeem, M.A. Al Ghamdi, M. Hussain, M.A. Khan, K.M. Khan, S.H. Almotiri, S.A. Butt, Brain Tumor Analysis Empowered with Deep Learning: A Review, Taxonomy, and Future  Challenges., Brain Sci. 10 (2020). https://doi.org/10.3390/brainsci10020118.

[11] Y. Zheng, Y. Zheng, D. Suehiro, S. Uchida, Top-rank convolutional neural network and its application to medical image-based diagnosis, Pattern Recognit. 120 (2021) 108138. https://doi.org/10.1016/j.patcog.2021.108138.

[12] D. Pan, A. Zeng, L. Jia, Y. Huang, T. Frizzell, X. Song, Early Detection of Alzheimer's Disease Using Magnetic Resonance Imaging: A Novel Approach Combining Convolutional Neural Networks and Ensemble Learning, Front. Neurosci. 14 (2020) 259. https://doi.org/10.3389/fnins.2020.00259.

[13] M. Liu, D. Cheng, W. Yan, A.D.N. Initiative, Classification of Alzheimer's Disease by Combination of Convolutional and Recurrent Neural Networks Using FDG-PET Images, Front. Neuroinform. 12

(2018) 35. https://doi.org/10.3389/fninf.2018.00035.

[14] J. Gu, Z. Wang, J. Kuen, L. Ma, A. Shahroudy, B. Shuai, T. Liu, X. Wang, G. Wang, J. Cai, T. Chen, Recent advances in convolutional neural networks, Pattern Recognit. 77 (2018) 354–377. https://doi.org/10.1016/j.patcog.2017.10.013.

[15] J.G. Lee, S. Jun, Y.W. Cho, H. Lee, G.B. Kim, J.B. Seo, N. Kim, Deep learning in medical imaging: General overview, Korean J. Radiol. 18 (2017) 570–584. https://doi.org/10.3348/kjr.2017.18.4.570.

[16] L. Aversano, M.L. Bernardi, M. Cimitile, R. Pecori, Deep neural networks ensemble to detect COVID-19 from CT scans, Pattern Recognit. 120 (2021) 108135. https://doi.org/10.1016/j.patcog.2021.108135.

[17] R. Rasti, M. Teshnehlab, S.L. Phung, Breast cancer diagnosis in DCE-MRI using mixture ensemble of convolutional neural networks, Pattern Recognit. 72 (2017) 381–390. https://doi.org/10.1016/j.patcog.2017.08.004.

[18] A.Z. Abidin, B. Deng, A.M. DSouza, M.B. Nagarajan, P. Coan, A. Wismüller, Deep transfer learning for characterizing chondrocyte patterns in phase contrast X-Ray computed tomography images of the human patellar cartilage, Comput. Biol. Med. 95 (2018) 24–33. https://doi.org/10.1016/j.compbiomed.2018.01.008.

[19] P. Qin, K. Wu, Y. Hu, J. Zeng, X. Chai, Diagnosis of Benign and Malignant Thyroid Nodules Using Combined Conventional Ultrasound and Ultrasound Elasticity Imaging, IEEE J. Biomed. Heal. Informatics. 24 (2020) 1028–1036. https://doi.org/10.1109/JBHI.2019.2950994.

[20] M.A. Morid, A. Borjali, G. Del Fiol, A scoping review of transfer learning

research on medical image analysis using ImageNet, Comput. Biol. Med. 128 (2020) 104115. https://doi.org/10.1016/j.compbiomed.2020.104115.

[21] X. Li, Y. Grandvalet, F. Davoine, A baseline regularization scheme for transfer learning with convolutional neural networks, Pattern Recognit. 98 (2020) 107049. https://doi.org/10.1016/j.patcog.2019.107049.

[22] J. Yosinski, J. Clune, Y. Bengio, H. Lipson, How transferable are features in deep neural networks?, Adv. Neural Inf. Process. Syst. 4 (2014) 3320–3328.

[23] V. Cheplygina, M. de Bruijne, J.P.W. Pluim, Not-so-supervised: A survey of semi-supervised, multi-instance, and transfer learning in medical image analysis, Med. Image Anal. 54 (2019) 280–296. https://doi.org/10.1016/j.media.2019.03.009.

[24] J. Song, Y.J. Chai, H. Masuoka, S.W. Park, S.J. Kim, J.Y. Choi, H.J. Kong, K.E. Lee, J. Lee, N. Kwak, K.H. Yi, A. Miyauchi, Ultrasound image analysis using deep learning algorithm for the diagnosis of thyroid nodules, Medicine (Baltimore). 98 (2019) e15133. https://doi.org/10.1097/MD.0000000000015133.

[25] Z. Zhu, E. Albadawy, A. Saha, J. Zhang, M.R. Harowicz, M.A. Mazurowski, Deep learning for identifying radiogenomic associations in breast cancer, Comput. Biol. Med. 109 (2019) 85–90. https://doi.org/10.1016/j.compbiomed.2019.04.018.

[26] P. Qin, K. Wu, Y. Hu, J. Zeng, X. Chai, Diagnosis of Benign and Malignant Thyroid Nodules Using Combined Conventional Ultrasound and Ultrasound Elasticity Imaging., IEEE J. Biomed. Heal. Informatics. 24 (2020) 1028–1036. https://doi.org/10.1109/JBHI.2019.2950994.

[27] M. Hu, K. Sim, J.H. Zhou, X. Jiang, C. Guan, Brain MRI-based 3D Convolutional Neural Networks for Classification of Schizophrenia and Controls., Annu. Int. Conf. IEEE Eng. Med. Biol. Soc. IEEE Eng. Med. Biol. Soc. Annu. Int. Conf. 2020 (2020) 1742–1745. https://doi.org/10.1109/EMBC44109.2020.9176610.

[28] J. Oh, B.L. Oh, K.U. Lee, J.H. Chae, K. Yun, Identifying Schizophrenia Using Structural MRI With a Deep Learning Algorithm, Front. Psychiatry. 11 (2020) 16. https://doi.org/10.3389/fpsyt.2020.00016.

[29] M. Hu, X. Qian, S. Liu, A.J. Koh, K. Sim, X. Jiang, C. Guan, J.H. Zhou, Structural and diffusion MRI based schizophrenia classification using 2D pretrained and 3D naive Convolutional Neural Networks, Schizophr. Res. (2021).

[30] C. Szegedy, S. Ioffe, V. Vanhoucke, A.A. Alemi, Inception-v4, inception-ResNet and the impact of residual connections on learning, 31st AAAI Conf. Artif. Intell. AAAI 2017. (2017) 4278–4284.

[31] L. Wang, A. Kogan, D. Cobia, K. Alpert, A. Kolasny, M.I. Miller, D. Marcus, Northwestern University Schizophrenia Data and Software Tool (NUSDAST), Front. Neuroinform. 7 (2013) 1–13. https://doi.org/10.3389/fninf.2013.00025.

[32] A. Kogan, K. Alpert, J.L. Ambite, D.S. Marcus, L. Wang, Northwestern University schizophrenia data sharing for SchizConnect: A longitudinal dataset for large-scale integration, Neuroimage. 124 (2016) 1196–1201. https://doi.org/10.1016/j.neuroimage.2015.06.030.

[33] N.F. Ho, J.E. Iglesias, M.Y. Sum, C.N. Kuswanto, Y.Y. Sitoh, J. De Souza, Z. Hong, B. Fischl, J.L. Roffman, J. Zhou, Progression from selective to general involvement of hippocampal subfields in

schizophrenia, Mol. Psychiatry. 22 (2017) 142–152. https://doi.org/10.1038/mp.2016.4.

[34] N.F. Ho, Z. Li, F. Ji, M. Wang, C.N. Kuswanto, M.Y. Sum, H.Y. Tng, Y.Y. Sitoh, K. Sim, J. Zhou, Hemispheric lateralization abnormalities of the white matter microstructure in patients with schizophrenia and bipolar disorder, J. Psychiatry Neurosci. 42 (2017) 242–251. https://doi.org/10.1503/jpn.160090.

[35] F. Kurth, C. Gaser, E. Luders, A 12-step user guide for analyzing voxel-wise gray matter asymmetries in statistical parametric mapping (SPM), Nat. Protoc. 10 (2015) 293–304. https://doi.org/10.1038/nprot.2015.014.

[36] J.L.R. Andersson, M. Jenkinson, S. Smith, Non-linear registration aka spatial normalisation, FMRIB Tech. Rep. TRO7JA2. 22 (2007) 22.

[37] B. Rashid, V. Calhoun, Towards a brain-based predictome of mental illness, Hum. Brain Mapp. 1–68 (2020). https://doi.org/10.1002/hbm.25013.